\documentclass[12pt]{article}

\usepackage{graphicx} 
\usepackage[affil-it]{authblk}
\begin{document}

\title{The role of CRKL in Breast Cancer Metastasis: Insights from Systems Biology}
\author{Abderrahim Chafik}
\affil{ Institut Mohamed VI, Sit\'e de l'air, 120101, T\'emara, Rabat, Morocco}

\maketitle

\abstract{MicroRNAs (miRNAs) are small non-coding RNAs that regulate gene expression post-transcriptionally. They are involved in key biological processes and then may play a major role in the development of human diseases including cancer, in particular their involvement in breast cancer metastasis has been confirmed. Recently, the authors of ref.(\cite{key1} have found that miR-429 may have a role in the inhibition of breast cancer metastasis and have identified its target gene CRKL as a potential candidate. In this paper, by using systems biology tools we have shown that CRKL is involved in positive regulation of ERK1/2 signaling pathway and contribute to the regulation of LYN through a topological generalization of feed forward loop}


\newpage 

\section{Introduction}
MicroRNAs are an evolutionarily conserved class of small (about 22 nucleotides) non-coding RNAs that negatively regulate gene expression post transcriptionally by base-pairing to the 3'UTR of target mRNAs leading to repression of protein production or mRNA degradation. Each miRNA can regulate the expression of hundreds of genes and  a single transcript can be targeted by multiple miRNAs. (For a recent introduction see for example \cite{key2} 

Since their discovery in 1993 \cite{key3} there have been increasing evidences that they are involved in important biological processes. Indeed, in normal cells, miRNAs control normal rates of cellular growth, proliferation, differentiation and apoptosis. Involvement of miRNAs in diverse cellular events indicates that dysregulated expression and function may lead to development  of differnt deseases including cancer. A potential role of  miRNAs in the development and progression of cancer has been suggested by some observations early in the history of miRNAs. one of these observations is that tumor cell lines and malignant tumors were found to have widespread deregulation of miRNA expression compared to normal tissues \cite{key4}. Thus, tumor formation may be a consequence of a reduction of a tumor suppressor miRNA or overexpression of an oncogene miRNA .Examples of tumor suppressor miRNAs are let-7 family, miR-29, miR-34 and miR -15, and of oncogene miRNAs are miR-17-92, miR-155 and miR-221.

miRNA involvement has been observed in almost every type of cancer including breast cancer \cite{key5, key6}. Breast cancer is the most common type of cancer among women. However, the majority of deths from breast cancer are not due to the primary tumor itself, but are the result of metastasis. Breast cancer metastasis is a very complex process that comprises a series of sequential steps and involves several processes \cite{key7}.

Several miRNAs have been found to be involved in the breast cancer metastasis. For instance, an up-regulated miR-21 down-regulates the PTEN which inhibits cell migration, spreading and focal adhesion. miR-10b which is an important miRNA is involved in causing epithelial-mesenchymal transition (EMT). 

MicroRNAs which are involved in activating the epithelial differentiation in breast cancer cells include the members of miR-200 family \cite{key4, key5}.The miRNA miR-429 belongs to this family that contains also miR-200a, miR-200b, miR-200c and miR-141. These five members are found in two clusters. In human, miR-200a, miR-200b and miR-429 are located on chromosome 1, while miR-200c and miR-141 are on chromosome 12. They are highly enriched in epithelial tissues \cite{key8}.

There are growing evidences that miR-200 microRNAs are involved in cancer metastasis. Indeed, they have been shown to inhibit epithelial-mesenchymal transition (EMT) which is the initiating step of metastasis by maintaining the epithelial phenotype through direct targeting of transcriptional repressors of E-cadherin, ZEB1 and ZEB2 \cite{key9, key10, key11}. There
is also evidence that the miR-200 family is up-regulated in distal breast metastasis indicating that these miRNAs are important for colonization of metastatic breast cancer cells through induction of mesenchymal to epithelial transition \cite{key12}.

It has been shown that miR-429 (with other two miRNAs) can be a potential biomarker of early human non-small cell lung cancer (NSCLC) \cite{key13}. It has also been demonstrated that the expression of miR-429 is often upregulated in NSCLC compared with normal lung tissues. It has been found that overexpression of miR-429 promotes cell proliferation, migration and invasion \cite{key14}. 

miR-429 was also found to be involved in pancreatic cancer stem cells \cite{key15}, colorectal cancer\cite{key16}, ovarian cancer \cite{key17} and also in breast cancer.
Recently a study carried out by Zhi-Bin Ye and coworkers revealed that decreased expression  of miR-429 are probably involved in negatively regulating bone metastasis of breast cancer cells \cite{key1}. On the other hand, its overexpression remarkably suppressed invasion. By combining the in silico analysis of miR-429 targets  and global transcriptional profile they have identified CRKL as a potential target that was probably a fundamental node in controlling bone metastasis of breast cancer. 

Crk-like (CRKL) is an adapter protein that has crucial roles in multiple biological processes, including cell proliferation, adhesion, and migration. It has been found to be overexpressed in many malignant tumors and plays crucial roles in tumor progression. It contributes to pancreatic cancer cell proliferation and invasion through ERK signaling \cite{key18}, and promotes lung cancer cell invasion through ERK-MMP9 pathway \cite{key19}. CrkL plays also a regulatory role in the SDF-1-induced Erk1/2 and PI3K/Akt pathways and further managed the invasion and migration of breast cancer cells \cite{key20}. It has been shown that overexpression of CRKL correlated with progression and malignant proliferation of human breast cancers \cite{key21}.

The purpose of this paper is perform a systems biology study to elucidate the role of CRKL in breast cancer metastasis and find out some of the biological processes in which it is involved. To achieve this goal we use systems biology tools.

\section{Materials and methods}

We begin by downloading the list of target genes of miR-429 from the DIANA-TarBase v7,0 database that catalogues published experimentally validated miRNA gene interactions \cite{key22}. With the list of genes we have just obtained we use IntAct to find curated interactions \footnote{IntAct is one of the largest available repositories for curated molecular interactions data, storing PPIs as well as interactions involving other molecules. It is hosted by the European Bioinformatics Institute. IntAct has evolved into a multisource
curation platform and many other databases curate into IntAct and make their data available through it \cite{key23}. } . 

To visualize the resulting network we use the cytoscape 3,2,0, an open source software platform for visualizing complex networks and integrating these with any type of attribute data. \cite{key24}.  In order to identify groups of genes that share a similar function we use the  clusterMaker, a Cytoscape application developed at UCSF that allows the user to easily create and visualize topological clusters using a great variety of methodologies, here we use the GLay Community Clustering algorithm.

In order to determine which Gene Ontology (GO) terms are significantly overrepresented in this set of genes we use the BiNGO Plugin of Cytoscape  \cite{key25}.

\section{Results and discussion}

\begin{figure}[h]
\centering
\includegraphics[width=.4\textwidth]{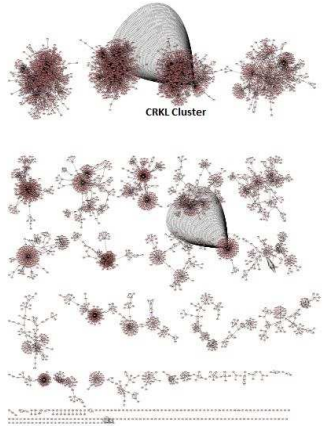}
\caption{Figure 1: The result of the Glay Community Clustering Algorithm applied to the targetome of miR-429}
\label{fig:clustered}
\end{figure}

By uploading the list of target genes onto cytoscape and after filtering by taxonomy identifier and choosing the human case the resulting network consists of 5172 nodes and 18177 edges. Then, by applying the clusterMaker cytoscape plugin we obtaing the result shown in figure figure~\ref{fig:clustered}

In our study we focus on the signaling process because of its important role in mediating each stage of tumour metastasis \cite{key26}. We apply the BINGO plugin to the cluster containing CRKL which consists of 700 nodes and 1192 edges. From the network view of the result of GO analysis by BINGO we see that all the branches going out from the regulation of signaling process term converge on the positive regulation of ERK 1 and ERK 2 cascade term which has a p-value equal to $4.5\times 10^{-10}$. The parent term is the positive regulation of MAPKKK cascade with a p-value equal to $2\times 10^{-16}$.

The extracellular signal-regulated kinase ERK 1/2 mitogen activated protein MAP kinase module is known as a conserved signaling pathway that plays a major role in the control of cell proliferation, survival and differentiation. Its involvement in cancer has been proved by convincing evidences. In vitro studies of cancer cell lines and in vivo findings from animal models have highlighted the role of ERK 1/2 signaling in cell survival, motility, invasiveness and angiogenesis. Accumulating evidences suggest that the ERK1/2 signaling pathway also contributes to the increased motility, invasiveness and dissemination of tumor
cells. ERK 1/2 may then contribute to every step of the metastatic process \cite{key26}. 

On another side, by performing a network motif search using visant, the analysis tool for biological networks and pathways\footnote{http://visant.bu.edu/}  we found that CRKL is involved in many feed forward loops, some of them contain LYN a member of Src family kinases.

\begin{eqnarray}
CRKL & --> & KHDRBS1 ->> LYN \\
CRKL & --> & GAB1 ->> LYN \\
CRKL & --> & ABI1 ->> LYN \\
CRKL & --> & EGFR ->> LYN \\
CRKL & --> & BCAR1 ->> LYN \\
CRKL & --> & MAP4K1 ->> LYN \\
CRKL & --> & SHC1 ->> LYN \\
CRKL & --> & SYK ->> LYN \\
CRKL & --> & TYK2 ->> LYN \\
CRKL & --> & CBL ->> LYN \\
CRKL & --> & KIT ->> LYN \\
CRKL & --> & BLNK ->> LYN 
\end{eqnarray}  

\begin{figure}[h]
\centering
\includegraphics[width=.4\textwidth]{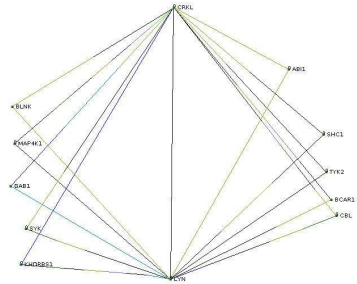}
\caption{Figure 2: A multi-y generalization of feed forward network motif where CRKL is the input and LYN is the output}
\label{ffl}
\end{figure}

Indeed, Src family kinases (SFK) such as Lyn mediate signal transduction and are involved in the regulation of tumor cell proliferation, adhesion, motility, and invasion, and contribute to tumor progression, metastasis, and angiogenesis . As a member of SFK, Lyn was found to play a role in inhibiting tumor growth and metastasis in Ewing's sarcoma \cite{key27}. It was also shown to be a mediator of epithelial-mesenchymal transition and target of dasatinib in breast cancer \cite{key28}. The functions of the Lyn tyrosine kinase in health and disease have been reviewed in \cite{key29}.

 According to figure \ref{ffl} CRKL and LYN are involved in a topological generalization of the feed forward loop (FFL) which consists in its basic form of only three nodes: an input (X), moderator (Y) and output (Z) node. It is encountered with some of its topological generalizations as a key functional component in gene transcription, signal transduction and neural networks \cite{key30}. However, the present case represents a multi-y generalization of the FFL which needs further study, in particular the determination of the sign of the different edges (activation or repression). We can suspect a combinatorial control of LYN by CRKL together with the other genes.
 
We have thus shown the involvement of CRKL in the positive regulation of ERK1/2 signaling pathway. We have also found that it may contribute to the regulation of LYN through a generalized FFL.

\bibliographystyle{apalike}
\bibliography{chafik}

\end{document}